\documentclass[%
 reprint,
 amsmath,amssymb,
 aps,
]{revtex4-1}

\usepackage{amsfonts}
\usepackage{bbold}
\usepackage{mathrsfs}
\usepackage{color}
\usepackage{ulem}
\usepackage{comment}
\usepackage{float}
\usepackage{graphicx}
\usepackage{dcolumn}
\usepackage{bm}

\begin{document}

\preprint{APS/123-QED}

\title{Universality class of the motility-induced critical point \\in large scale off-lattice simulations of active particles}

\author{Claudio Maggi$^{1,2}$}
 \email{claudio.maggi@roma1.infn.it}
\author{Matteo Paoluzzi$^{3}$}
\author{Andrea Crisanti$^{2,4}$}

\author{Emanuela Zaccarelli$^{4,2}$}
\author{Nicoletta Gnan$^{4,2}$}
\email{nicoletta.gnan@roma1.infn.it}

\affiliation{
$^1$NANOTEC-CNR, Institute of Nanotechnology, Soft and Living Matter Laboratory, Roma, Italy \\ 
$^2$ Dipartimento di Fisica, Sapienza Universit\`a di Roma, Piazzale A. Moro 2, I-00185, Rome, Italy \\
$^3$ Departament de Física de la Mat\`eria Condensada, Universitat de Barcelona, C. Martí Franqu\`es 1, 08028 Barcelona, Spain \\
$^4$ CNR-ISC, Institute of Complex Systems, Roma, Italy
\\}

\begin{abstract}

We perform large-scale computer simulations of an off-lattice two-dimensional model of active particles undergoing a motility-induced phase separation (MIPS) to investigate the systems critical behaviour close to the critical point of the MIPS curve. By sampling steady-state configurations for large system sizes and performing finite size scaling analysis we provide exhaustive evidence that the critical behaviour of this active system belongs to the Ising universality class.
In addition to the scaling observables that are also typical of passive systems, we study the critical behaviour of the kinetic temperature difference between the two active phases. This quantity, which is always zero in equilibrium,  displays instead a critical behavior in the active system which is well described by the same exponent of the order parameter in agreement with mean-field theory.


\end{abstract}

\maketitle

\paragraph*{Introduction.}

One of the pillars of statistical physics is the concept of universality in critical phenomena. In equilibrium systems, close to a second-order phase transition, universality can be ultimately attributed to the divergence of the correlation length of the order parameter.  The  behavior of this growing length-scale is found to be independent on the microscopic details of the systems but is determined only by few specific features, i.e. the spatial dimensionality and the symmetries of the order parameter as firstly hypothesized by Kadanoff~\cite{kadanoff1971critical}. Depending on such parameters it is possible to trace back the critical behaviour of disparate systems within few groups, called universality classes.

One of the biggest challenge of recent years is to transfer the vast knowledge acquired on universal behaviour of equilibrium systems into active matter physics. Active matter represents a peculiar class of non-equilibrium systems where the elementary units or agents are self-propelled objects capable to convert energy in systematic movement~\cite{Marchetti13,Bechinger17}. The interacting agents are often complex biological objects that exhibit self-organized behavior at large scales giving rise to many and diverse living materials~\cite{klopper2018physics}. 
In particular, self-propulsion can trigger a feedback between motility and local density causing an effective attractive interaction between active particles~\cite{Tailleur08}. This attractive force can bring to a phase separation in active fluids, remarkably similar to the gas-liquid coexistence in equilibrium systems, which is called Motility-Induced Phase Separation (MIPS)~\cite{cates2015motility}. Since MIPS is a very general feature
of active dynamics, observed independently on the details of the self-propulsion, it might play a role also in biological systems. For instance, it has been recently observed that multicellular aggregates of {\it Myxococcus xanthus} might take advantage of density-motility feedback for developing large scale collective behaviors that have been well described by MIPS~\cite{AdamPRL}.

\begin{figure*}[!t]
\includegraphics
[width=1.\textwidth]{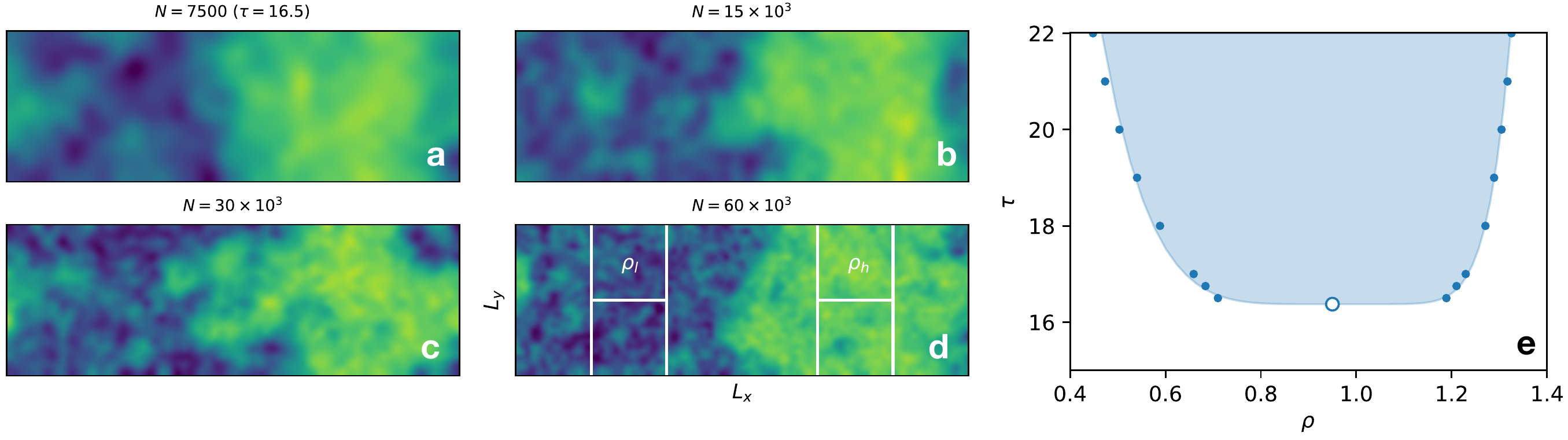}
\caption{
\footnotesize{(a)-(d)  {Smoothed} density field in a  {rectangular} geometry for
four active systems of different sizes at $\tau=16.5$ (where phase separation becomes appreciable). The color map encodes the local density value from yellow (high density) to blue (low density). The configurations have been shifted so that the dense and diluted phases are centered onto the four sub-boxes (panel (d)) considered in the analysis. The average density in the right sub-boxes is denoted by $\rho_h$ (the density of the high-density phase), while the average density of the left sub-boxes by $\rho_l$ (low-density). (e) Coexistence curve constructed with the densities of the dense and diluted phases (filled data-points), the estimated critical point is shown as an open circle.}
}
\label{fig:snap}
\end{figure*}

In equilibrium physics the standard gas-liquid coexistence ends with a critical point that belongs to the Ising universality class~\cite{callen1998thermodynamics,domb2000phase}. A natural question is whether or not MIPS curve ends in a critical point and if there is a region close to phase separation in which the active critical behaviour that can be traced back to a specific universality class.
Effective equilibrium approaches~\cite{paoluzzi2019statistical,Paoluzzi}, 
and field-theoretic computations~\cite{PhysRevLett.123.068002} have previously pointed to the Ising universality class. Concerning numerical simulations, despite numerous works have addressed the properties of phase-separated MIPS states \cite{stenhammar2014phase,PhysRevE.95.012601,Fily12,levis2017active,patch2018curvature,PhysRevE.100.052604,PhysRevLett.121.098003,mandal2019motility}, the study of the critical region and the determination of the critical properties still remain challenging and controversial. In particular it has been shown that active Brownian particles in two-dimensions (2$d$) display some critical exponents deviating considerably from the Ising ones~\cite{siebert2018critical}. However recent on-lattice simulations of an active model have shown that critical exponents are in good agreement with the Ising universality class~\cite{PhysRevLett.123.068002} and suggested that simulations done off-lattice have been performed far from the scaling regime due to the small size of the systems.
 {Since it is not \textit{a priori} clear if both models lie in the same universality class{,} 
one must rely on large-scale computer simulations which are often a necessary tool to understand possible discrepancies between off-lattice and on-lattice critical exponents. This has been the the case, for example, for equilibrium Heisenberg fluids ~\cite{nijmeijer1995monte,mryglod2001ferromagnetic}.}

In this work, we report results of large-scale off-lattice simulations of Active Ornstein-Uhlenbeck particles (AOUPs) at criticality. Performing Finite-Size-Scaling (FSS) analysis, we show that the system's critical exponents agree with the Ising universality class. We also show the kinetic temperature difference between the two phases, emerging at criticality, display a critical behavior which is well described by the exponent 
of the order parameter in agreement with mean-field theory combined with a small-$\tau$ expansion of the AOUPs model.

\paragraph*{Model and Methods.}
We consider a system composed of $N$ self-propelled AOUP disks  
in 2$d$~\cite{Maggi,Szamel15}. This model is perhaps the simplest active particle model (due to the linearity of the process producing the ``active noise'') which has lead to numerous novel theoretical developments~\cite{PhysRevLett.117.038103,dal2019linear,bonilla2019active}. It has been shown~\cite{PhysRevLett.117.038103} that AOUPs exhibit MIPS for large values of the persistence time of the activity, as also displayed in Fig.~\ref{fig:snap}(a)-(d).
The equations of motion of AOUPs read

\begin{align} \label{micro1}
\dot{\mathbf{r}}_i = \boldsymbol{\psi}_i + \mathbf{F}_i \\ \label{micro2} \tau \, \dot{\boldsymbol{\psi}}_i = -\boldsymbol{\psi}_i + \boldsymbol{\eta}_i
\end{align}

\noindent where we indicate with $\mathbf{r}_i$ the $i$-th particle's position, with $\boldsymbol{\psi}_i$ the self-propelling force, and with $\mathbf{F}_i = \sum_{j\neq i} \mathbf{f}_{ij}$ the conservative force acting on the particle. We consider two-body interactions, i.e., $\mathbf{f}_{ij} = -\nabla_{\mathbf{r}_i} \phi(r_{ij})$, with $r_{ij} = | \mathbf{r}_i - \mathbf{r}_j| $ and use a simple inverse power-law potential $\phi(r)=(r/\sigma)^{-12}/12$ with a cut-off at $r=2.5\,\sigma$. Here $\sigma$ represents the diameter of the particle and is set to 1.
In Eq.~(\ref{micro2}) $\tau$ is the persistence time of the active force and $\boldsymbol{\eta}$ is a standard white noise source, i.e. $\langle \eta_{i}^\alpha(t) \rangle=0$ and $\langle \eta_{i}^\alpha(t) \eta_{j}^\beta(s) \rangle = 2 D\, \delta_{ij} \delta^{\alpha \beta} \delta(t-s)$, where the greek indices indicate Cartesian components. Here $D$ is the diffusivity of the non-interacting particles which is related to the mean squared velocity $v$ by $D=v^2 \tau$. In all simulations we fix $v=1$ and increase $\tau$ from small to large values to observe the transition as shown by the schematic phase diagram in Fig.~\ref{fig:snap}(e).
Eq.s~(\ref{micro1}),(\ref{micro2}) have been integrated numerically using the Euler scheme with a time step $\Delta t=10^{-3}$ up to $N_t=10^9$ time steps for the largest systems which is adequate to observe full relaxation of the density auto-correlation function as shown in the Supplemental Material (SM). In addition we start each simulation from a random initial configuration and we perform up to $N_e=10^8$ equilibration steps which guarantees that all runs reach the steady-state, even close to the critical point. We perform averages over several independent runs and errors reported represent twice the standard error of the mean (see SM for details).

The active particles move in a rectangular box of size $L_x \times L_y$ with 1:3 ratio ($L_x=3 \, L_y$) and periodic boundary conditions. The simulated systems size are $N=(7.5,15,30,60)\times 10^3$. 
To avoid spurious effects due to the presence of an interface~\cite{rovere1993simulation,siebert2018critical} we compute the quantities of interest only in four sub-boxes of size $L=L_y/2$ centered on the dense and diluted phases. These boxes are located at $x=L_x/2\pm L_x/4$ and at $y=L_y/2\pm L_y/4$. To ensure that these positions coincide with the locations of the dense and diluted phases, as in ~\cite{siebert2018critical}, for each configuration we first find the system center of mass along $x$ (with periodic boundaries~\cite{bai2008calculating}), and we shift all particles so that the $x$-coordinate of the center of mass coincides with $x=L_x/2+L_x/4$ as shown in Fig.~\ref{fig:snap}(d). 
 {
We stress that the method also works at the critical point if the two phases are distinguishable enough. This is typically the case when the critical order parameter distribution shows two peaks as in many Ising~\cite{plascak2013probability} and Lennard-Jones systems~\cite{potoff1998critical}.
This technique has been successfully applied to the $2d$ Ising model~\cite{siebert2018critical} and to the lattice active model~\cite{PhysRevLett.123.068002}.
}

All simulations are performed at fixed density $\rho=0.95$ by varying accordingly the box size $L_x = 3 L_y = (3 N/\rho)^{1/2}$. 
{ {This value approximately corresponds to the critical density estimated for the smallest investigated system size, which is found to be $\rho_c = 0.953(0.037)$ (errors from fit are given in brackets, see SM for details). }}
%

\paragraph*{Results.}

\begin{figure*}[!t]
\includegraphics
[width=1.\textwidth]{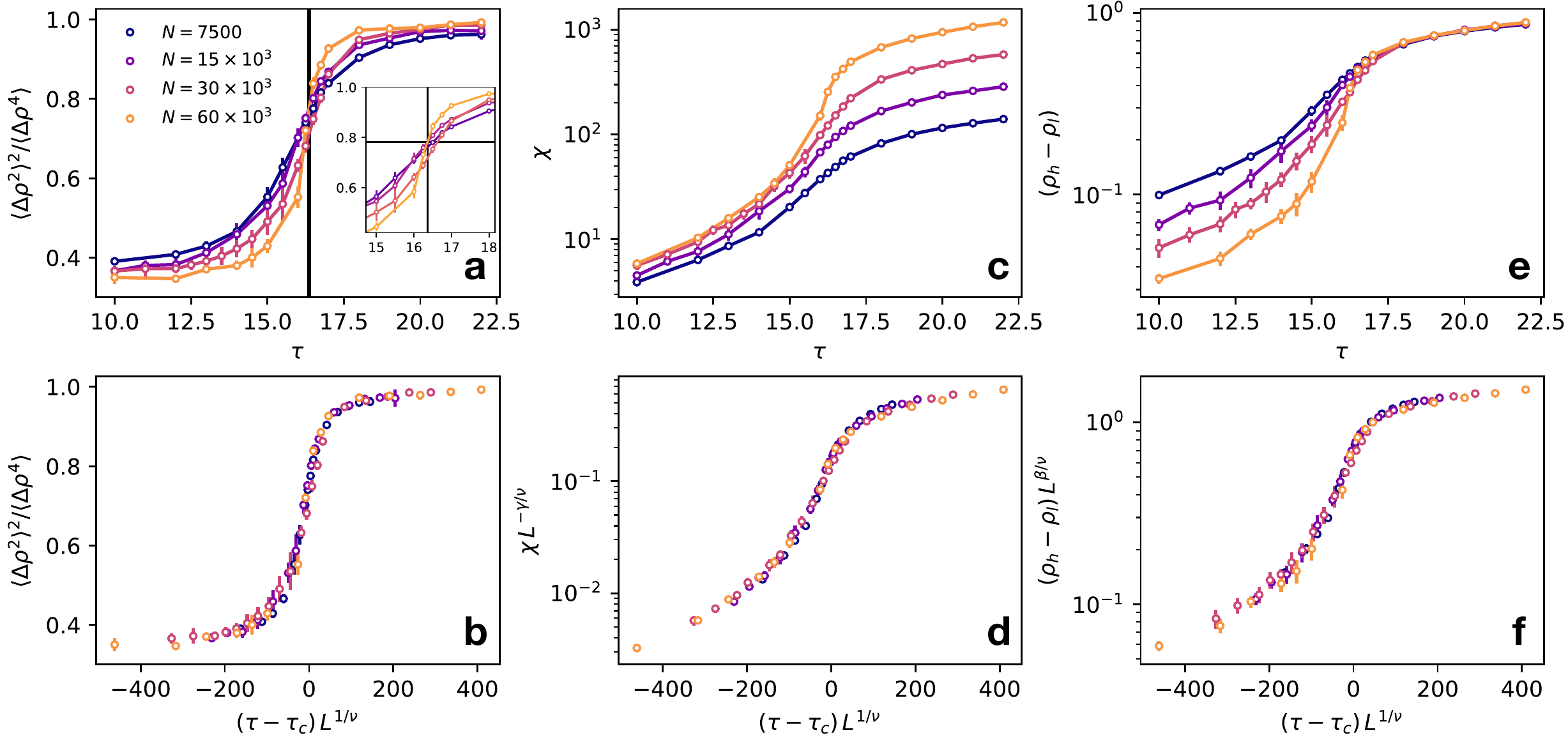}
\caption{\footnotesize{FSS analysis. (a) Binder parameter for different system sizes. The intersection of the curves allows to locate $\tau_c$ (vertical line). 
 {The inset shows a magnification in a small region close to $\tau_c$ (i.e. $|\tau-\tau_c|/\tau_c<0.1$), {where} the straight black lines indicate the crossing point of $N=15\times 10^3$ and $N=60\times 10^3$ determining the critical point}; (b) Collapse of data in (a) as a function of the scaling variable $(\tau-\tau_c)L^{1/\nu}$ with $\nu=1$; 
(c) susceptibility $\chi$ as a function of $\tau$ for different system sizes; (d) collapse of data in (c) onto a universal scaling function with the exponents $\gamma=7/4$ and $\nu=1$;  (e) order parameter  for different system sizes; (f) data collapse of data in (e) with the exponents $\beta=1/8$ and $\nu=1$. The color code is the same for all panels (see legend in (a)).}
}
\label{binder}
\end{figure*}


Although finite systems can not develop any diverging correlation length, the finite-size scaling hypothesis allows us to systematically study the critical properties away from the thermodynamic limit~\cite{amit2005field}. Using the finite-size scaling ansatz, we assume that a generic observable $\mathcal{O}$ near the critical point behaves as 
$\mathcal{O} = L^{\frac{\zeta_\mathcal{O}}{\nu}} \left[ F_\mathcal{O}(L \xi^{-1}) + O(L^{-\omega},\xi^{-\omega})\right]$,
where $\zeta_\mathcal{O}$ is the critical exponent associated with the observable $\mathcal{O}$, $F_\mathcal{O}$ is a universal finite-size scaling function and $\omega$ is the power of the (subleading) correction-to-scaling exponent~\cite{amit2005field}. Here $\nu$ is the exponent associated with the divergence of the correlation length $\xi$ as the control parameter is varied across the transition. In our active particle system the relaxation time of the noise $\tau$ is the control parameter, therefore we assume $\xi \sim (\tau-\tau_c)^{-\nu}$. Using this and ignoring sub-leading corrections we get $\mathcal{O}=L^{\zeta_\mathcal{O}/\nu} \, G_\mathcal{O}(L^{1/\nu} (\tau-\tau_c))$ (where $G_\mathcal{O}$ is a universal scaling function). This implies that, if the correct $\tau_c$, $\nu$ and $\zeta_\mathcal{O}$ are known, all values of $\mathcal{O}$ measured for different sizes should collapse onto each other when $L^{-\zeta_\mathcal{O}/\nu} \mathcal{O}$ is plotted as a function of $L^{1/\nu} (\tau-\tau_c)$.

A particularly interesting observable is the fourth order cumulant of density fluctuations ${\langle \Delta \rho^2 \rangle^2/\langle \Delta \rho^4 \rangle}$ (the Binder parameter~\cite{binder1981finite,rovere1988block,rovere1990gas}), where brackets indicate averages over configurations and over sub-boxes.
The density fluctuations are computed in the four $L\times L$ sub-boxes described above, specifically $\langle \Delta \rho^2 \rangle = \langle (N_b/L^2-\langle N_b/L^2 \rangle)^2 \rangle$ where $N_b$ is the number of particles found in one single sub-box.
For the Binder parameter we expect $\zeta_{\mathcal{O}}=0$ and thus it should be size-independent at $\tau=\tau_c$. 
 {Exploiting this property we locate $\tau_c=16.361(0.058)$ and $\mathcal{B}=[{\langle \Delta \rho^2 \rangle^2/\langle \Delta \rho^4 \rangle}]_{\tau=\tau_c}=0.781(0.017)$ as the intersection of the data for $N=15\times 10^3$ and $N=60\times 10^3$ (Fig.~\ref{binder}(a)). We choose to use these two sizes because $N=60\times 10^3$ is the largest {simulated size and $N=15\times 10^3$ is two times smaller in linear size. 
The estimated value of $\mathcal{B}=0.781(0.017)$ is lower than the corresponding value found in the triangular lattice gas ($\mathcal{B}=0.8321(0.0023)$, see SM), but it is close to that found in the active lattice model~\cite{PhysRevLett.123.068002} ($\mathcal{B} \approx 0.75$).}
Note that $\tau_c=16.36$ is approximately the value at which cumulants of all sizes cross as shown in the inset of Fig.~\ref{binder}(a) where we report a magnification of the main panel in a small $\tau$-interval around $\tau_c$ (see SM for a systematic study of the crossing points). Fig.~\ref{binder}(b) shows a good data collapse of the cumulant data-points with the Ising exponent $\nu=1$. A direct way~\cite{siebert2018critical} to determine $\nu$ is to consider the size dependence of the slope of the cumulants at $\tau=\tau_c$, this method yields $\nu = 1.03(0.10)$ as shown in the SM.}

Next we test the scaling of the susceptibility $\chi = \langle (N_b-\langle N_b \rangle)^2 \rangle/\langle N_b \rangle$ (shown in Fig.~\ref{binder}(b)). Fig.~\ref{binder}(c) shows that scaling is very good using the Ising critical exponent $\gamma= 7/4$. 
 {In the SM we also show that, if we fit directly the size-dependent values of $\chi$ at $\tau_c$, we obtain $\gamma=1.84(0.20)$ which is compatible with the Ising $\gamma$. }
Note also that the $\chi$ in Fig.~\ref{binder}(c) does not show the typical peaked shape of the Ising model. This is due to the fact that the $\chi$ is obtained here by averaging together the values of $N_b$ both in the dense and diluted phase. In the SM we show that the $\chi$ (computed in the same way) for the lattice gas display a similar s-shaped curve as a function of the inverse temperature and that it scales with $\gamma=7/4$.
Furthermore, in Fig.~\ref{binder}(e) we consider the density difference between the boxes centered in the high and low-density  phases $(\rho_h-\rho_l)$ (see Fig.~\ref{fig:snap}(d)), which corresponds to the order parameter of the system. We find that this quantity displays a good scaling with an exponent Ising $\beta=1/8$ (Fig.~\ref{binder}(f)). It is worth to stress that, in the thermodynamic limit, the order parameter would be different from zero only for $\tau>\tau_c$. It is however expected that, for finite systems, a smooth variation of the order parameter should be found also below $\tau_c$ and that this should scale with the appropriate exponent. We have also checked that $(\rho_h-\rho_l)$, computed as in the active system, scales with $\beta=1/8$ in the case of the 2$d$ equilibrium lattice gas for temperatures above the critical temperature (see SM for details).  
 {A direct fit of the size-dependent critical $(\rho_h-\rho_l)$ gives $\beta=0.113(0.055)$, which is compatible with the Ising $\beta=0.125$. To improve the accuracy {of the $\beta$ estimate} we apply a finer technique finding the value of the exponent which optimizes the data collapse (see SM). This yields $\beta= 0.133(0.022)$.}
 
 \begin{figure}[!th]
\includegraphics
[width=0.9\columnwidth]{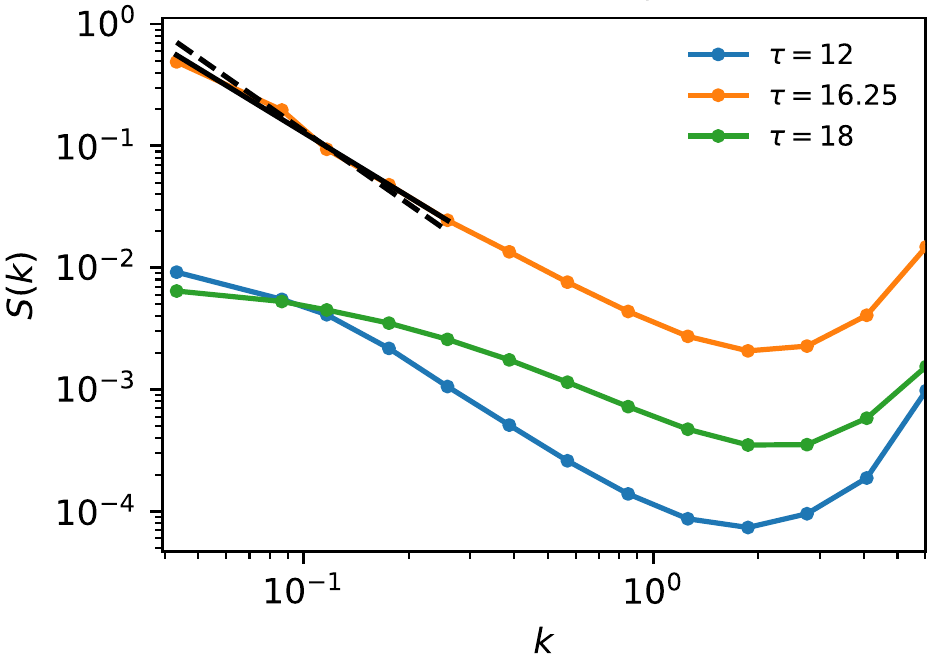}
\caption{Static structure factor
$S(k)$ computed in the dilute phase for the largest system ($N=60\times 10^3$), different 
colors refer to different values of $\tau$ (see legend). Approaching the critical point the structure factor is well fitted by a power law $S(k)\sim k^{-2+\eta}$ at low $k$ with $\eta=1/4$ (full line). The best fit with $S(k)\sim k^{-2}$ (mean-field) is also shown as a dashed line for comparison. }
\label{sk}
\end{figure}

\begin{figure}[!th]
\includegraphics
[width=0.99\columnwidth]{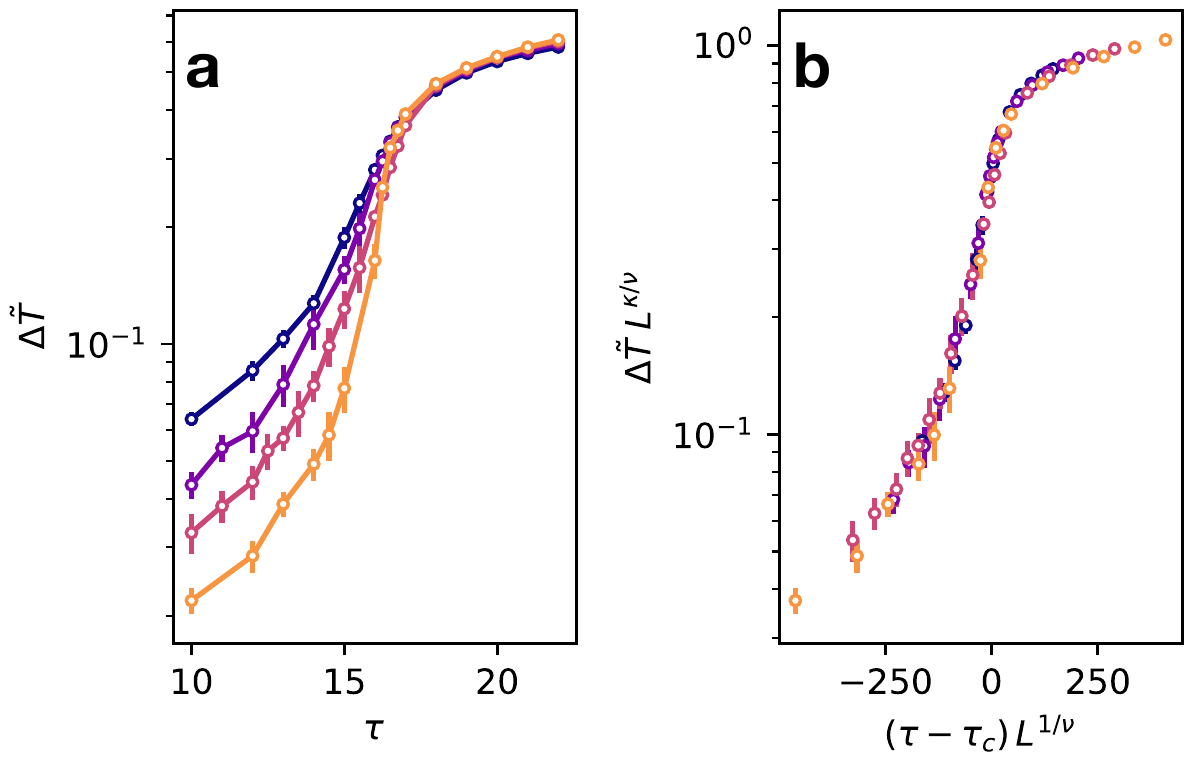}
\caption{ (a) Difference between the average squared particle velocity of the diluted and dense phases plotted as a function of $\tau$ (different colors indicate different system sizes, same legend as Fig.~2(a)). (b) Data of (a) scaled with the exponents $\nu=1$ and $\kappa=\beta=1/8$. }
\label{velsca}
\end{figure}

Next we check if our results are consistent with the Ising exponent $\eta=1/4$ which controls the decay of the static structure factor $S(k)$ near the critical point, i.e. $S(k) \sim k^{-2+\eta}$. To this aim we compute 
$S(k)=A \, \langle \rho_{\mathbf{k}}^\ast\, \rho_{\mathbf{k}} \rangle $ where $\rho_\mathbf{k}$ is the Fourier transform of the density fluctuations and the normalization factor $A$ is chosen so that $S(0)=\langle \Delta N^2 \rangle/\langle N \rangle$ ($N$ being the fluctuating number of particles in the sub-sytem considered). To avoid the interfaces and focus only onto the bulk phase we compute the $S(k)$ for the particles in the diluted phase considering only those particles having $|x-L_x/4|<L/2$, i.e. all particle in the left sub-boxes in Fig.~\ref{fig:snap}(d). We show the resulting $S(k)$ in Fig.~\ref{sk}: close to criticality $S(k)$ becomes fairly linear in double log scale at low $k$. The data at low $k$ are well fitted by the power law $k^{-2+\eta}$ with $\eta=1/4$ (full line) which is appreciably different from the mean-field decay $S(k)\sim k^{-2}$ (dashed line).  { A direct power-law fit of these points gives $2-\eta = 1.709(0.090)$ and 
$\eta=0.290(0.090)$ which are compatible with the Ising values $2-\eta = 1.75$ and $\eta=0.25$}.

{
Up to this point we have discussed {quantities} which display a critical behavior also in equilibrium fluids. We show now an {observable}  that is zero in equilibrium while it exhibits a singular behavior in the active case. Since in active systems the instantaneous velocities are coupled to positions~\cite{marconi2016velocity,PhysRevLett.117.038103}, whenever MIPS occurs dense regions of slow particles coexist with dilute regions of fast ones~\cite{mandal2019motility}. We then consider the average squared speed of particles in the dense and dilute sub-boxes that we indicate, respectively, with $\langle |\dot{\mathbf{r}}|^2 \rangle_h$ and $\langle |\dot{\mathbf{r}}|^2 \rangle_l$. The quantity 
$\Delta \tilde{T} = \frac{1}{2}(\langle |\dot{\mathbf{r}}|^2 \rangle_l-\langle |\dot{\mathbf{r}}|^2 \rangle_h)$ can be seen as the (effective) kinetic temperature difference between the two phases and its behavior it is shown in Fig.~\ref{velsca}(a). As expected $\Delta \tilde{T}$ decreases, as the two phases progressively mix upon lowering $\tau$, suggesting a scaling $\Delta \tilde{T} \sim (\tau-\tau_c)^\kappa$ with $\kappa>0$. More interestingly, $\Delta \tilde{T}$ shows a clear size dependence and we find a good data collapse if we use the exponents $\kappa=\beta=1/8$ and $\nu=1$ (Fig.~\ref{velsca}(b)). A direct estimate of the exponent gives $\kappa=0.122(0.022)$ satisfying $\kappa=\beta$ within the errors (see SM). Moreover we have been able to  theoretically derive the relation $\kappa=\beta$, within mean-field theory, by using a small-$\tau$ approximation of the AOUP model as shown in the SM.
}

\paragraph*{Discussion and Conclusions.} 
In this article, we have studied the critical properties of an active system undergoing MIPS in two spatial dimensions. Performing large-scale numerical simulations on GPU we have demonstrated that the critical behavior of the system agrees well with the Ising universality class. 
 {
It is worth to stress the importance of simulating large system sizes: previous studies on off-lattice active models have reported different values of the critical exponents~\cite{siebert2018critical}. Although it has been speculated~\cite{PhysRevLett.123.068002} that the limited sizes employed did not allow to observe the scaling regime, it is true that similar sizes have been exploited for the study of critical passive attractive liquids, finding numerical results compatible with the Ising universality class~\cite{rovere1993simulation}.
We instead suspect that for those sizes another correlation length, different from the critical one, {may interfere} with the scaling behaviour of the active system.  A very recent work~\cite{caporusso2020micro} has shown that the dense phase formed by active particles undergoing MIPS is made of a mosaic of hexatic micro-domains. We find that (see SM for discussion), already at the critical point, the hexatic correlation length is comparable with the size of the sub-boxes employed for the FSS analysis when the system size is small ($N=3750$) justifying the choice of larger system sizes. Indeed for a size as small as $N=3750$ we find that the crossing point of the Binder cumulant happens at quite low values, although a reasonable scaling is found also for this size (see SM). }

 {Our large-scale simulation results are also consistent with recent works taking into account non-integrable active terms in a {field}-theoretical framework~\cite{caballero2018bulk}. These results indicate that, when full MIPS is possible, these extra terms are irrelevant in a renormalization group sense and the system belongs to the Ising universality class. On the other hand, far from criticality, these non-equilibrium contributions could produce significant differences with respect to an equilibrium gas-liquid phase separation~\cite{wittkowski2014scalar,nardini2017entropy,PhysRevLett.123.148005,PhysRevX.8.031080,mandal2019motility}. Within this context it would be interesting to understand how one could make the active critical point unstable~\cite{caballero2018bulk} by altering the microscopic interactions and/or the dynamics.\\
}


\paragraph*{Acknowledgments.} EZ and NG acknowledge financial support from the European Research Council (ERC Consolidator Grant 681597, MIMIC). MP acknowledges financial support from the H2020 program and from the Secretary of Universities and Research of the Government of Catalonia through Beatriu de Pin\'os program Grant No. 2018 BP 00088.

\bibliography{mpbib.bib}

\clearpage

\newpage

\onecolumngrid

\renewcommand{\thefigure}{S\arabic{figure}}
\renewcommand{\theequation}{S\arabic{equation}}

\setcounter{figure}{0}

\section*{Supplemental Material}
\subsection*{Equilibration time and length of the simulation runs}

\begin{figure*}[h]
\includegraphics
[width=\textwidth]{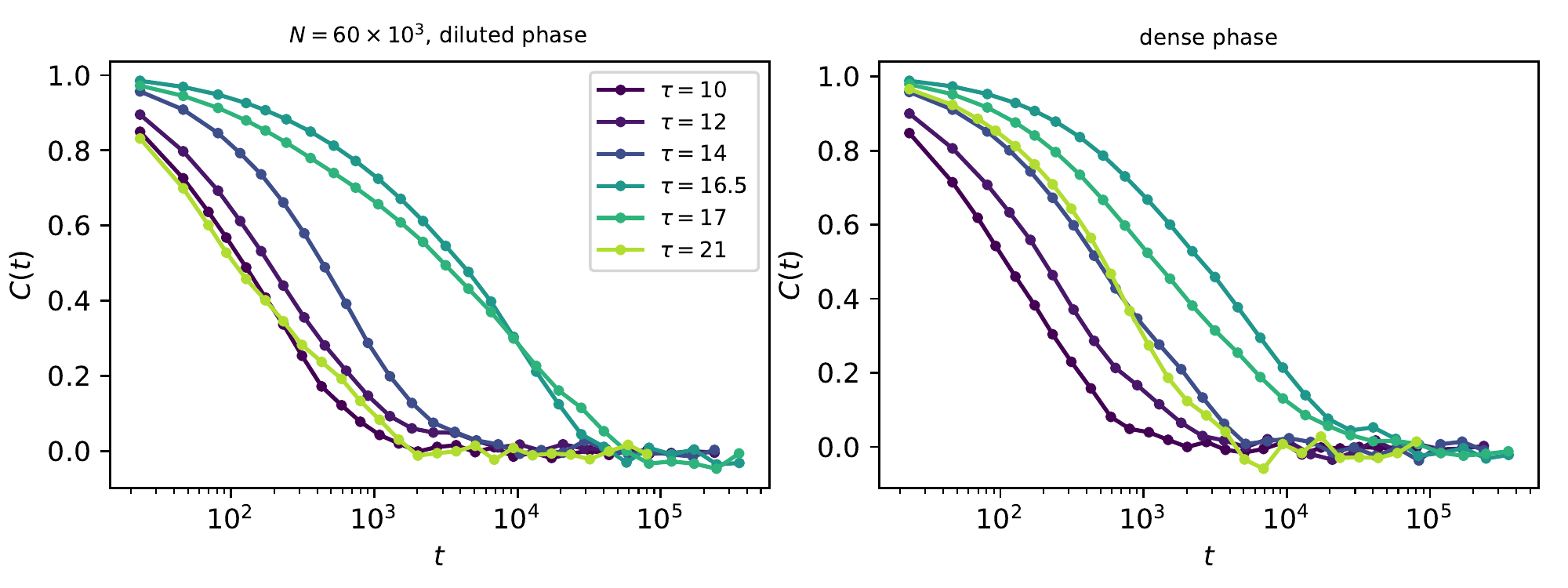}
\caption{Density auto-correlation function for the system with $N=60\times 10^3$ particles at $\rho=0.95$ and different values of $\tau$. Left panel: diluted phase. Right panel: dense phase. }
\label{fig:corrfun}
\end{figure*}

We show here that the length of the simulation runs is long enough to allow the full relaxation of the density correlation function $C(t)=\langle \Delta N_b(0) \Delta N_b(t) \rangle/\langle \Delta {N_b}^2\rangle$, where $\Delta N_b=N_b - \langle N_b \rangle$ is the fluctuation of the number of particles in the sub-box. More specifically we analyze separately the relaxation in the dense and diluted phases considering the two sub-boxes on the left and two on the right as shown in Fig~1 of the main text. 
Figure~\ref{fig:corrfun}(a) and (b) display $C(t)$ for the largest system investigate (i.e $N=60\times 10^3$) at different $\tau$ values both in the dense and diluted phase: in both cases $C(t)$ decay to zero at all $\tau$.

\subsection*{Averages and error estimation}

All quantities appearing in the main text (i.e. the Binder cumulant, the susceptibility and the order parameter) are averaged over $16000$ to $36000$  configurations and over all sub-boxes. Individual configurations are taken at time intervals of duration $\approx23$ in reduced units, which is approximately equal the largest $\tau$ explored. Moreover for the values of $\tau$ close to $\tau_c$ these quantities of interest are  averaged over multiple initial random configurations (up to 12 for the largest system sizes).

To estimate the error on these quantities we proceed as follows: we first divide each run in time windows larger than the relaxation time of the density correlation function $C(t)$ (Fig.~\ref{fig:corrfun}) we than compute the observable in each time-window and compute the standard error of the mean over all time windows. In Fig.~2 of the main text we report the error as twice the standard error of the mean.
We have also checked that by computing the average of the Binder cumulant over these time windows (instead that on all configurations) we get almost identical results than those reported in Fig.~2 of the main text.

\subsection*{Estimate of the critical $\rho$ and $\tau$}

\begin{figure}[!h]
\includegraphics
[width=0.99\columnwidth]{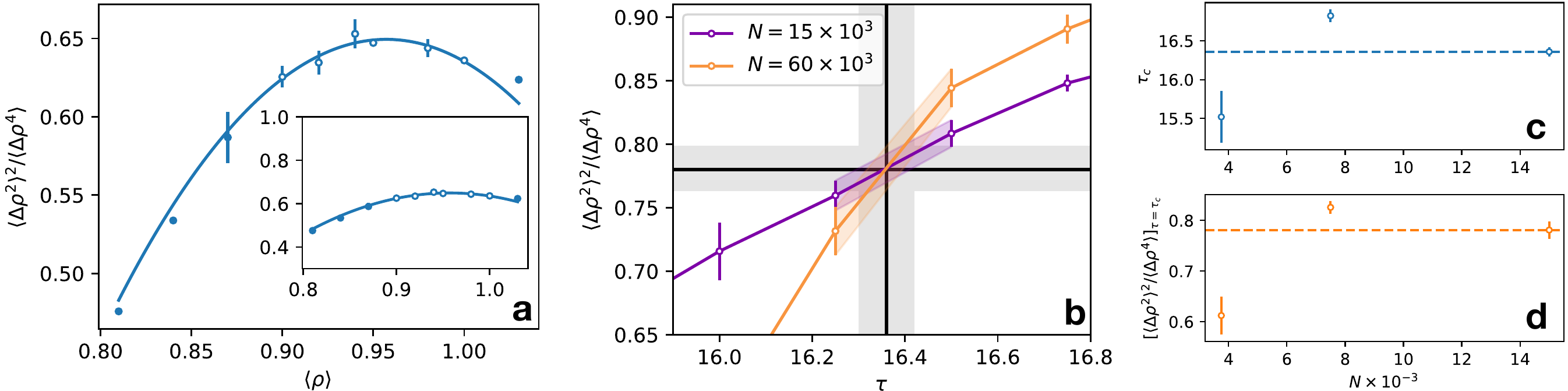}
\caption{(a) Fourth order cumulant as a function of density. Data points represent the Binder parameter varying the density at fixed $\tau=16$ and $N=3750$. The open symbols are the closest six points to the maximum used in the parabolic fit (full line) for determining $\rho_c$. The inset shows the same data of the main panel plotted on the same $y$-range of Fig.~2(a) of the main text. (b) Intersection of the cumulants (colored points) for $N=60 \times 10^3$ and $N=15\times 10^3$ used for locating $\tau_c$. The intersection point $\tau_c=16.361(0.058)$ and $\mathcal{B}=0.781(0.017)$ (black lines) is found by a piece-wise interpolation of the cumulant curves. The error on $\tau_c$ and $\mathcal{B}$ (gray areas) is obtained by propagating the $y$-error on the points nearby the intersection (colored areas). (c) and (d) (same $x$-axis) show the obtained values of $\tau_c$ and $\mathcal{B}$ (open symbols) as a function of the system size. The values of the reference point for the largest size is also reported as a dashed line.}
\label{fig:rho_scan}
\end{figure}

 {We roughly estimate the critical density by performing a density scan, at fixed $\tau$, in the proximity of the critical point for the smallest system investigated (i.e. $N=3750$).
According to Ref.~\cite{rovere1990gas} the cumulant should exhibit a maximum at $\rho=\rho_c$ when plotted as a function of $\rho$ and at fixed $\tau = \tau_c$.
Fig.~\ref{fig:rho_scan}(a) shows that the Binder parameter  {indeed displays} a maximum if we fix $\tau=16$. This value has been chosen based on preliminary simulations and it is close to the critical value $\tau_c=16.36$ estimated in the following. Note also that the cumulant varies much less upon changing density in this small interval than upon changing $\tau$ on a large interval as in Fig.~2(a) of the main text. This is shown in the inset of Fig.~\ref{fig:rho_scan}(a) plotting the same data of the main panel in the $y$-range $(0.3,1)$, i.e. the range of the binder cumulant as a function of $\tau$.
To extract $\rho_c$ we have fitted with a 2nd order polynomial the six closest points to the maximum in Fig.~\ref{fig:rho_scan}(a) finding $\rho_c = 0.953(0.037)$ where the fit error is reported in brackets}. We set the value $\rho_c=0.95$ for all the sizes discussed in the main text neglecting the dependence of $\rho_c$ on the size of the system. 

 {
The critical value $\tau$, indicated by $\tau_c$, has been determined by finding the crossing of the cumulants for two sizes $N=15\times 10^3$ and $N=60\times 10^3$. These have been chosen because $N=60\times 10^3$ is the largest size simulated and $N=15\times 10^3$ has a linear size which is two times smaller than the largest one. The procedure followed to find the intersection is illustrated in Fig.~\ref{fig:rho_scan}(b). We linearly interpolate the cumulant curves and find $\tau_c$ as the $x$-intersection of {the} two lines, {while} the critical cumulant value $\mathcal{B}=[\langle \Delta \rho^2\rangle^2/\langle \Delta \rho^4 \rangle]_{\tau=\tau_c}$ {is found from the} intersection on the $y$-axis.  We obtain $\tau_c=16.361(0.058)$ and $\mathcal{B}=0.781(0.017)$. This $\mathcal{B}$ value is lower than the one {found for} the triangular lattice gas ($\mathcal{B}=0.8321(0.0023)$, see section below) and {for} the square lattice gas ($\mathcal{B}\approx 0.83$ extracted from Ref.~\cite{siebert2018critical}). However it is close to $\mathcal{B} \approx 0.75$  which is the critical cumulant value of the active lattice model found in Ref.~\cite{PhysRevLett.123.068002}. Note that previous studies on the Lennard-Jones fluid have also reported a lower value of the critical Binder parameter with respect to the Ising 
model~\cite{rovere1993simulation}.\\
\indent To further check the size dependence {of} $\tau_c$ and $\mathcal{B}$,  we have repeated this procedure also {considering other system sizes, that are separated by a factor of 2 in linear size,} i.e.: ${(N=3750,N=15\times 10^3)}$ and ${(N=7500,N=30\times 10^3)}$. The resulting $\tau_c(N)$ and $\mathcal{B}(N)$ are shown in Fig.~\ref{fig:rho_scan}(c) and (d) respectively where each value of $\tau_c(N)$ and $\mathcal{B}(N)$ is associated with the smallest size in the pair. In Fig.~\ref{fig:rho_scan}(c) and (d) we see that the values of both quantities
at $N=7500$ are closer {to those of the largest system size (dashed lines) than to those corresponding to $N=3750$}.
This suggests that, upon increasing the size, $\tau_c(N)$ and $\mathcal{B}(N)$ progressively converge to the infinite-system critical values.
}

\subsection*{Direct estimates of the critical exponents}

Here we show how we directly estimate the critical exponents. Following Ref.~\cite{siebert2018critical} we first focus on the dependence of the slope of the critical Binder cumulant on $L$ whose scaling with size is controlled by the exponent $\nu$:

\begin{equation} \label{slope}
\left[
\frac{\partial}{\partial \tau}
\left(
\frac{
\langle \Delta \rho^2 \rangle^2}
{\langle \Delta \rho^4 \rangle}
\right)
\right]_{\tau=\tau_c} \sim L^{1/\nu}
\end{equation}

To use Eq.~(\ref{slope}) we evaluate the derivative of the cumulant by fitting the numerical data with a generalized logistic function of the form

\begin{equation}
y(\tau) = A_1+\frac{A_2}{[A_3+A_4 \,  e^{-(\tau-\tau_0)}]^{1/\theta}}
\end{equation}

\begin{figure*}[!h]
\includegraphics
[width=1.\textwidth]{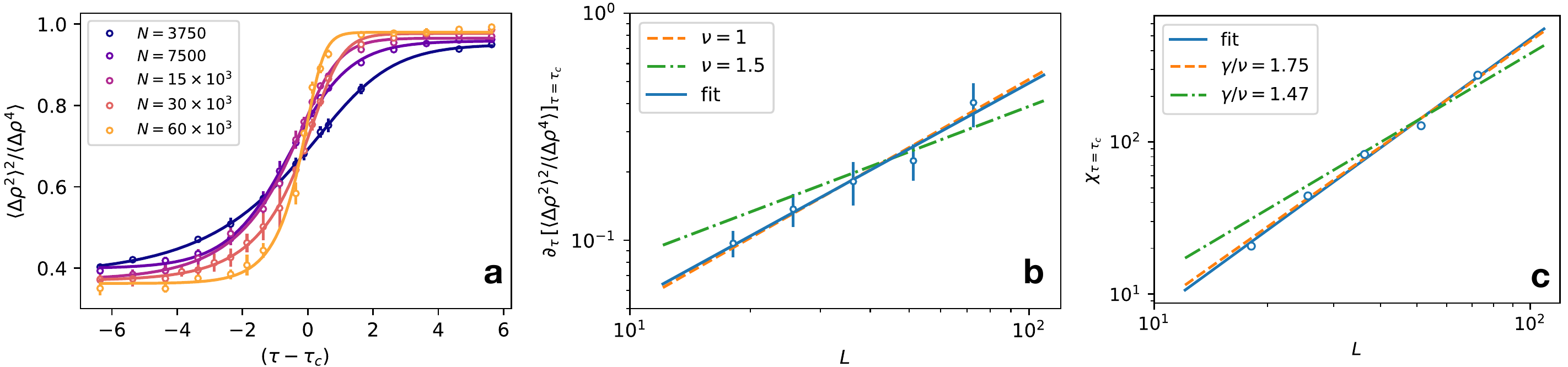}
\caption{(a) Fits of the Binder cumulant for estimating its derivative near $\tau_c$. (b) Slope of the cumulants as a function of size. Solid line is the best fit of data points which gives $\nu=1.03$. Orange dashed line is the best fit fixing  $\nu=1$ while green dashed-dotted line is the best fit with $\nu=1.5$ from Ref.~\cite{siebert2018critical}. (c) susceptibility as a function of size. Solid line is the best fit of data points which gives $\gamma/\nu=1.787$. Orange dashed line is the best fit fixing $\gamma/\nu=1.75$ while the green dashed-dotted line is the fit with $\gamma/\nu=1.47$ taken from Ref.~\cite{siebert2018critical}. }
\label{fig:slope}
\end{figure*}

Where $A_1,A_2, A_3,A_4,\tau_0$ and $\theta$ are fitting parameters. As shown in Fig.~\ref{fig:slope}(a) this function fits well the data especially around $\tau_c$ and allows us to estimate the derivative (\ref{slope}). This derivative is reported in Fig.~\ref{fig:slope}(b)  as a function of the system size and it is indeed well fitted by $\sim L^{1/\nu}$ with $\nu=1$. 
 {In fact a direct fit with a power law gives $1/\nu = 0.968(0.096)$ (the fit error is indicated in brackets) and, by linear error propagation, $\nu = 1.03(0.10)$}.
Contrarily the best fit with the exponent $\nu=1.5$ proposed in Ref.~\cite{siebert2018critical} deviates considerably from the data.

Next we consider the size dependence of the susceptibility $\chi$ at $\tau_c$ (i.e. $\chi_{\tau=\tau_c}$), that should scale as $\chi_{\tau=\tau_c} \sim L^{\gamma/\nu}$. To do this we simply linearly interpolate the $\chi$ at $\tau_c$ at all sizes and report the results in Fig.~\ref{fig:slope}(c). This quantity is also well fitted by the Ising exponent $\gamma/\nu = 7/4 =1.75$.  {A direct fit with a power law yields $\gamma/\nu = 1.787(0.090) $. Using $\nu$ found above and propagating also its error we get $\gamma = 1.84(0.20)$.}
Also in this case by fixing the values of $\gamma = 2.2$ and $\tau = 1.5$ (i.e. $\gamma/\nu \approx 1.47$) from Ref.~\cite{siebert2018critical} we obtain a worse fit of our data.


\begin{figure*}[!h]
\includegraphics
[width=1.\textwidth]{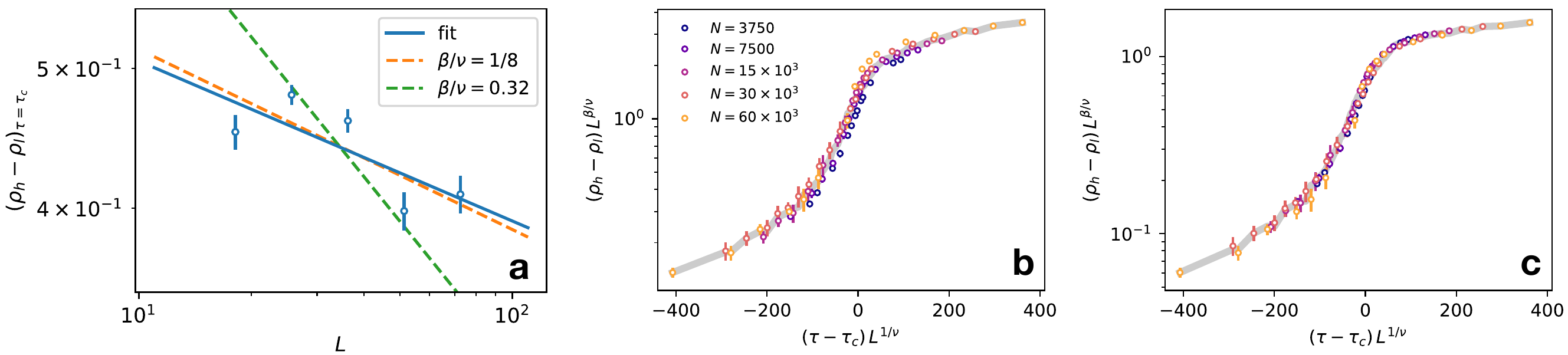}
\caption{ (a) Order parameter as a function of size at $\tau=\tau_c$. Solid line is the best fit of data points which gives $\beta/\nu=0.110(0.053)$ (i.e. $\beta=0.113(0.055)$). Orange dashed line is the best fit fixing $\beta/\nu=0.125$ while the green dashed-dotted line is the fit with $\beta/\nu=0.32$ taken from Ref.~\cite{siebert2018critical}. 
(b) Collapsed order parameter data-points (colored dots) on the interpolating function (gray thick line) with exponent $\beta/\nu= 0.32$ from \cite{siebert2018critical}, the distance of the points from the interpolating function is large resulting in a large $\mathcal{E}$.
(c) Same as (b) but with the fitted parameter $\beta/\nu=0.130$ minimizing the $\mathcal{E}$ function.}
\label{fig:beta}
\end{figure*}

 {
To directly estimate $\beta$ we interpolate the points of the order parameter $(\rho_h-\rho_l)$ at $\tau_c$ for all sizes and we plot them as a function of $L$ in Fig.~\ref{fig:beta}(a). It is evident that these points are quite noisy  and also that $\beta/\nu$ is small. However we are still able to obtain an estimate of $\beta$ compatible with the Ising value ($\beta=0.125$) when we fit these points with a power law $(\rho_h-\rho_l)\sim L^{-\beta/\nu}$. This yields $\beta/\nu=0.110(0.053)$  (full line in Fig.~\ref{fig:beta}(a)) and $\beta=0.113(0.055)$ if the error on $\nu$ is propagated linearly. Note that this is close to the Ising value as shown by the orange line in Fig.~\ref{fig:beta}(a) and appreciably smaller than the $\beta=0.45$ of Ref.~\cite{siebert2018critical} shown by the green line Fig.~\ref{fig:beta}(a).\\
\indent
To obtain a more accurate estimate of $\beta$ we also implement a finer method which finds the exponent by minimizing the deviation between collapsed data. This type of technique has been applied in the past to extract the critical exponents of various spin models~\cite{bhattacharjee2001measure,houdayer2004low}. To practically apply this method we fix the values of $\tau_c$ and $\nu$ to the values determined above ($\tau_c=16.36$ and $\nu=1.03$). We then consider the following error function to be minimized:
\begin{equation} \label{eq:chi2_eq}
\mathcal{E} = \sum_i \left[
L_i^{\beta/\nu} m(\tilde{\tau_i},L_i)-
G(\tilde{\tau_i})
\right]^2
\end{equation}
where $L_i$ and $\tilde{\tau}_i={L_i}^{1/\nu}(\tau_i-\tau_c)$ are respectively the system size and the scaled control parameter of the $i$-th data-point, while  $m(\tilde{\tau_i},L_i)=[\rho_h - \rho_l]_{(\tilde{\tau}=\tilde{\tau}_i,\,L=L_i)}$ is the order parameter value at $L_i$ and $\tilde{\tau}_i$ (the sum runs over all available data). The function $G$ in Eq.~(\ref{eq:chi2_eq}) is the scaling function describing the critical behavior of $m$ whose analytic form is unknown. To circumvent this problem we evaluate the function $G$ by interpolating the values of $L^{\beta/\nu}m(\tilde{\tau},L)$ with a smooth function. We compute $G$ by averaging over windows of fixed size $\Delta \tilde{\tau}$ which we choose to be 10 times smaller than the overall $\tilde{\tau}$ range, i.e. $\Delta \tilde{\tau} = \max(|\tilde{\tau}_i|)/10$. In this way a smoothed $G$ can be evaluated at each desired value  $\tilde{\tau}_i$. An example of the resulting $G$ is plotted in Fig.~\ref{fig:beta}(b) where we use the parameter $\beta/\nu=0.32$ of Ref.~\cite{siebert2018critical}. It is clear that, while the resulting $G$ is smooth enough, the simulation data-points do not collapse well on the curve. The value of $\beta/\nu$ which minimizes the $\mathcal{E}$ function of Eq.~(\ref{eq:chi2_eq}) is $\beta/\nu=0.130(0.018)$ that results in a good data collapse as shown in Fig.~\ref{fig:beta}(c) and is again compatible with the Ising value $\beta/\nu=0.125$. This gives 
$\beta=0.133(0.022)$, by linear error propagation, which is also compatible with the Ising value.
}

\begin{figure*}[!h]
\includegraphics
[width=0.8\textwidth]{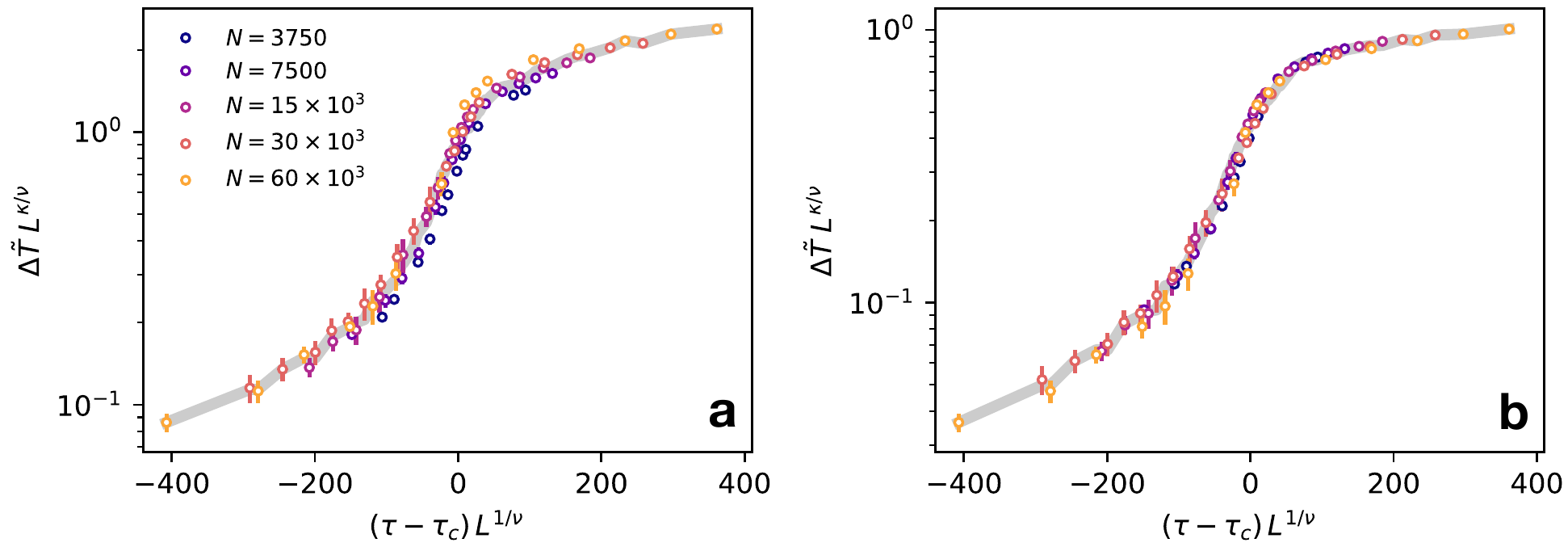}
\caption{ (a) Colored dots represent the values of 
$\Delta \tilde{T}$ collapsed on the interpolating function (gray thick line) with exponent $\kappa/\nu= 0.32$. (c) Same as (b) but with the fitted parameter $\kappa/\nu=0.118$ minimizing the $\mathcal{E}$ function.}
\label{fig:kappa}
\end{figure*}

Finally we estimate the exponent characterizing the critical behavior of the difference between the particle average squared speed of the dilute and dense phases, i.e. the quantity: $\Delta \tilde{T} = \frac{1}{2}(\langle |\dot{\mathbf{r}}|^2 \rangle_l-\langle |\dot{\mathbf{r}}|^2 \rangle_h)$. We use again the collapse optimization method introduced above for estimating $\beta$ (see Eq.~(\ref{eq:chi2_eq})).
In Fig.~\ref{fig:kappa}(a) we report the $\Delta \tilde{T}$ values if we use the exponent $\kappa/\nu=0.32$. It is evident that, with this exponent, the data-points do not collapse well on the interpolating function (gray curve in Fig.~\ref{fig:kappa}(a)). Minimizing the $\mathcal{E}$-function we obtain 
$\kappa/\nu =0.118(0.018)$ (i.e. $\kappa=0.122(0.022)$) which gives a good data collapse as shown in Fig.~\ref{fig:kappa}(b). 

\subsection*{Derivation of the exponent identity $\kappa=\beta$}

 {
To derive the exponent identity $\kappa=\beta$ we consider an AOU particle in $d=1$ and subjected to an external potential $\Phi(x)$. It is known~\cite{marconi2016velocity,martin2020statistical} that for small $\tau$ the velocity distribution of this particle is is a zero-centered Gaussian with variance 
\begin{equation} \label{varv2}
\langle {\dot{x}}^2 \rangle = v^2-v^2 \tau \,  \Phi''(x) + \mathcal{O}(\tau^2) 
\end{equation}
\noindent to first order in $\tau$. Here $v^2=D/\tau$ is the free particle mean squared velocity (which is assumed to be kept constant as in our simulations) and ${\Phi''(x)=\partial_{x^2} \Phi(x)}$ is the potential curvature. We now assume that the total potential curvature $\Phi$ felt by the probe particle in $x$ is generated by the interactions with other particles: ${\Phi''(x)=\sum_i\phi''(x-x_i)}$,
 where ${\phi''(x-x_i)}$ is the second derivative of the pair interaction potential. This can be rewritten as ${\Phi''(x)=\int dx' \, \hat{\rho}(x-x') \phi''(x-x')}$ where the integral extends over all space and we have introduced the density field $\hat{\rho}(x)=\sum_i\delta(x-x_i)$. 
 By ignoring density fluctuations (mean-field approximation) we set ${\hat{\rho}(x)=\rho=\mathrm{const}}$ and we obtain $\Phi''(x)=\overline{\phi}_2 \, \rho$, where $\overline{\phi}_2=\int dx' \phi''(x-x')$ the mean potential curvature which is assumed to be positive. 
 By using this in Eq.~(\ref{varv2}) and neglecting higher order corrections we get:
  $\langle {\dot{x}}^2 \rangle = v^2(1-\tau \, \overline{\phi}_2 \rho) $. We now consider the difference between the averaged squared speed in the low and high density phases: ${\Delta \tilde{T}=\frac{1}{2}(\langle {\dot{x}}^2 \rangle_l-\langle {\dot{x}}^2 \rangle_h) = \frac{1}{2} v^2\tau\overline{\phi}_2 (\rho_h-\rho_l)}$. If we now assume that $(\rho_h-\rho_l)\sim (\tau-\tau_c)^\beta$ near the critical point we have:
\begin{equation}
\Delta \tilde{T} \sim (\tau-\tau_c)^\kappa
\end{equation}
\noindent with $\kappa=\beta$, which is the relation verified by the simulation data within errors.
The derivation of this identity can be easily generalized to higher dimensions leading to the same result.
}

\subsection*{Scaling behavior of the susceptibility and of the order parameter in the 2$d$ equilibrium lattice gas}

\begin{figure*}[!h]
\includegraphics
[width=0.8\textwidth]{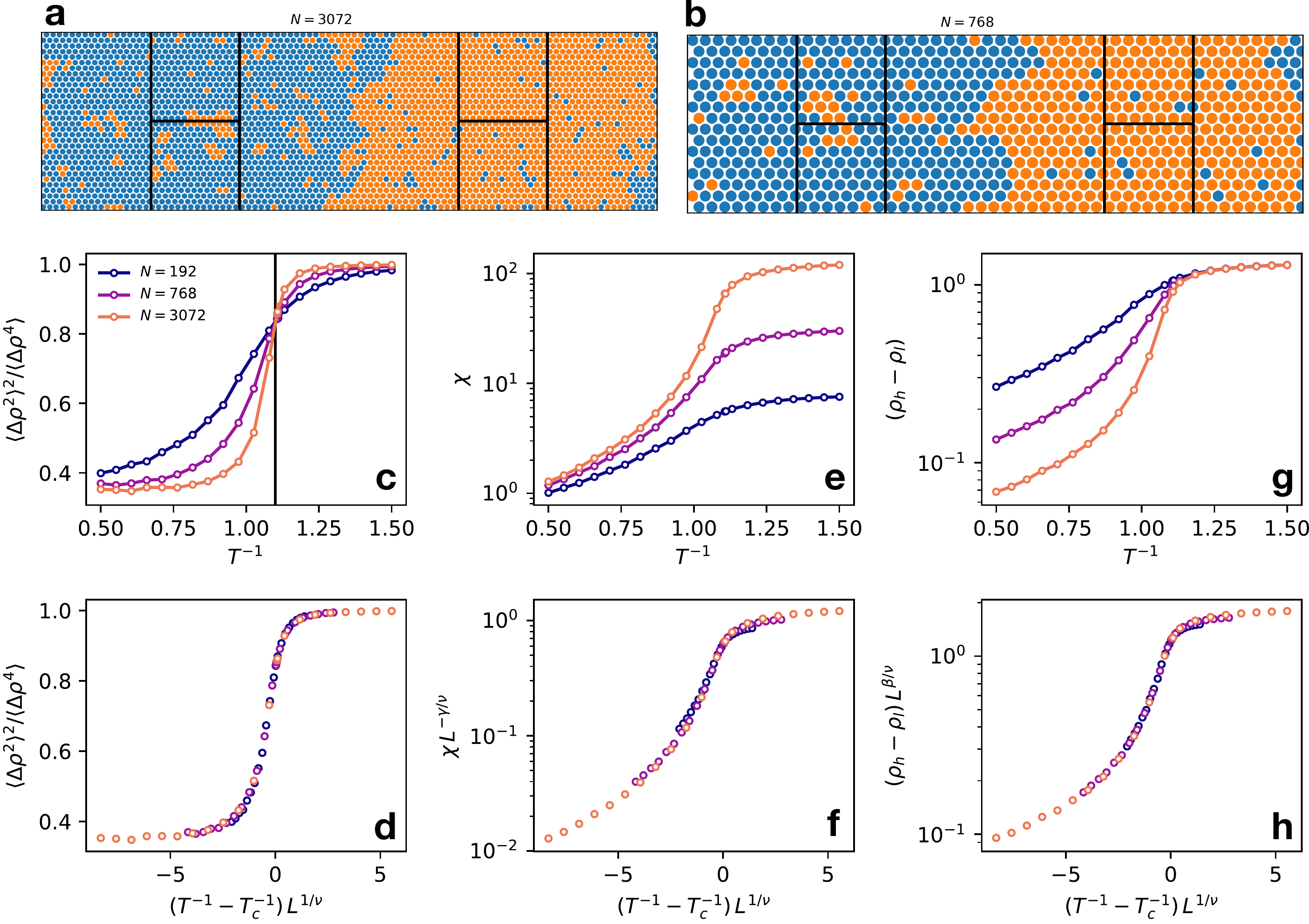}
\caption{(a) and (b) Near-critical configurations of an equilibrium lattice gas on the triangular lattice in a rectangular geometry for two different sizes (at $T^{-1} \approx 1.13$). Blue points represent empty sites while orange points represent occupied sites. The configuration is shifted so that the dense phase is centered on the right boxes (located at $x=3L_x/4$, black lines) and the diluted phase is centered on the left sub-boxes (at $x=L_x/4$). 
(c) Binder cumulant for the lattice gas as a function of inverse temperature for different sizes (see legend), the vertical line corresponds to the exact $T_c^{-1}\approx1.1$.. (d) Data collapse of the data in (c) with $\nu=1$.
(e) Susceptibility for the lattice gas as a function of inverse temperature for different sizes. (f) Data collapse of the data in (e) with $\nu=1$ and $\gamma=7/4$.
(g) Order parameter as a function of the inverse temperature for the lattice gas for various sizes indicated in the legend (the order parameter is computed as the average density difference between the sub-boxes on the right and on the left). (h) Data collapse of the data in (h) using the exponents $\beta=1/8$ and $\nu=1$.}
\label{ising}
\end{figure*}

To check whether the susceptibility and the order parameter behave in the same qualitative way in the active system and in equilibrium  we consider a lattice gas on a triangular lattice in a rectangular geometry (similar to the one employed for the active system). We simulate systems of three different sizes composed by $N= 192$, $768$ and $3072$ sites.
These sites are enclosed in a rectangular box of size $(0, L_x) \times (0, L_y)$ 
with $L_x = a\,N_x$ and $L_y = a \sqrt{3} N_y/2$, where $a=1$ is the lattice spacing. In all simulations we set $N_x=3\,N_y$ and $N=N_x \times N_y$ is the total number of sites. Some near-critical configurations lattice gas simulated is shown in Fig.~\ref{ising}(a) and (b).
By imposing periodic boundary conditions every site has $6$ neighbours and the total lattice gas Hamiltonian ($H_\mathrm{lg}$) is given by

\begin{equation}
\label{eq:dynamics_MC}
H_\mathrm{lg} = - J \sum_{\langle i,j\rangle} n_i \, n_j \,,
\end{equation}

\noindent where $J$ is the coupling constant set to $1$ for convenience and $n_i$ is the occupancy of the $i$-th site which assume the values $0$ or $1$. The simulations conserve the total occupancy (i.e. $\sum_i n_i = \mathrm{const} $) by using a Kawasaki-type dynamics~\cite{bovier2015kawasaki} in which a site can exchange its occupancy with any other site in the lattice in order to accelerate the approach to equilibrium. After an occupancy switch is proposed a standard Monte Carlo (MC) Metropolis rule is applied and the new configuration is 
accepted or rejected according to the energy change.
All simulation results are obtained at fixed average occupancy $\sum_i n_i/N = 0.5$ (i.e. at the critical occupancy), starting from a random configuration (i.e. at infinite temperature). 
It is possible to show, via the transformation $n_i = (1+\sigma_i)/2$, that the model~\eqref{eq:dynamics_MC} can be mapped onto the Ising model with spin $\sigma_i=\pm 1$ on the triangular lattice having critical temperature $T_c=4/\ln 3 \approx 3.641$ (for $J=1$ and $k_B=1$)~\cite{zhi2009critical}.
As a consequence, the $T_c$ of the lattice gas model turns out to be $T_c = 1/(\ln 3) \approx 0.91$, i.e. an inverse critical temperature ${T_c}^{-1} \approx 1.099$ while the critical average occupancy is $n_c = 0.5$. 

The configurations of the lattice gas are analyzed as described in the main text for the active system. We start by shifting each configuration so that the its center of mass is positioned at $x=3L_x/4$ as also shown in Fig.~\ref{ising}(a) and (b). Subsequently the quantities of interest are averaged over all four $L \times L$ sub-boxes, where $L = L_y/2$. The density $\rho$ in one sub-box is computed as $\rho = \sum_i'n_i/L^2$, where the prime indicates the sum runs only on those sites within the sub-box.
Using this method we further check the correctness of the critical temperature by  showing the Binder parameter and its good scaling with $\nu=1$ in Fig.s~\ref{ising}(c) and (d). 
 {By interpolating and averaging the values of the cumulants at the known value of $T_c$ for all sizes we get $\mathcal{B}=0.8321(0.0023)$ which is close to the value of $\mathcal{B}$ of the square lattice gas found in Ref.~\cite{siebert2018critical}}. 
In the main text we have mentioned that the $\chi$, computed by averaging over all sub-boxes, does not show the typical peaked shape but rather forms a s-shaped curve when plotted as a function of the control parameter. This is the case also for the equilibrium lattice gas as shown in Fig.~\ref{ising}(e). In Fig.~\ref{ising}(f) we also show that this $\chi$ scales well with $\nu=1$ and $\gamma=7/4$.
In the main text we have also used the average difference of the density in the high-density phase $\rho_h$ and of the low-density phase $\rho_l$ as an order parameter. To check if this quantity behaves as expected at criticality also in the equilibrium case we compute $\rho_h$ and $\rho_l$ as the average density of the two sub-boxes on the right and on the left respectively. The resulting $(\rho_h - \rho_l)$ is shown in Fig.~\ref{ising}(b) as a function of $T^{-1}$. In Fig.~\ref{ising}(c) we show that we obtain a good data collapse by using the Ising exponents $\beta=1/8$ and $\nu=1$.
These data are clearly compatible with those presented for the off-lattice active system discussed in the main text, thus reinforcing the robustness of the analysis bringing to the Ising universality class in the case of the active system.

\subsection*{System sizes, hexatic order and velocity correlation length}

\begin{figure*}[!t]
\includegraphics
[width=1.\textwidth]{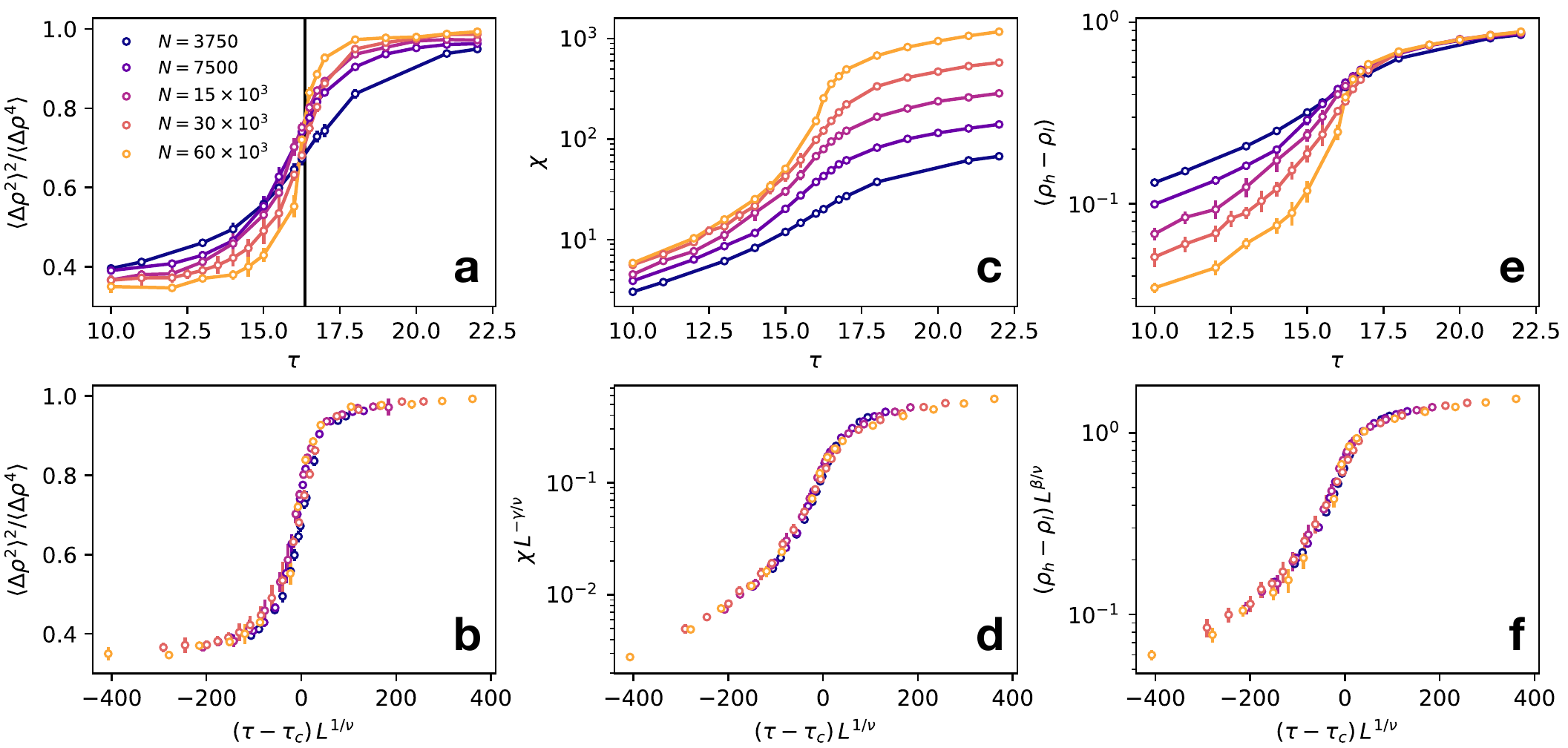}
\caption{Data collapse of the analyzed quantities with the exponents estimated directly $\nu=1.03$, $\gamma=1.84$ and $\beta=0.133$ (including the smallest system with size $N=3750$).
}
\label{binder_all}
\end{figure*}

We discuss here the data collapse for the smallest system simulated, i.e. $N=3750$ (not included in the main text), which is comparable with the sizes used in a previous investigation on the critical behaviour of an off-lattice  active system~\cite{siebert2018critical}. In Fig.~\ref{binder_all} we report the data for this size for the cumulant, the susceptibility and the order parameter. It is evident that a reasonable data collapse with the exponents calculated above is found also for $N=3750$ (see Fig.~\ref{binder_all}(b),(d) and (f)). However the  crossing point of the Binder cumulant for this size seems significantly lower in height and $\tau$ than the larger sizes (see also Fig.~\ref{fig:rho_scan}(c) and (d)). 

 {As mentioned in the main text we speculate that this could be due to the presence of another growing (but not diverging) correlation length. In the following we identify and compare two of them: the first related the hexatic order and the second associated to velocity correlations.
}

\begin{figure*}[!]
\includegraphics
[width=0.975\textwidth]{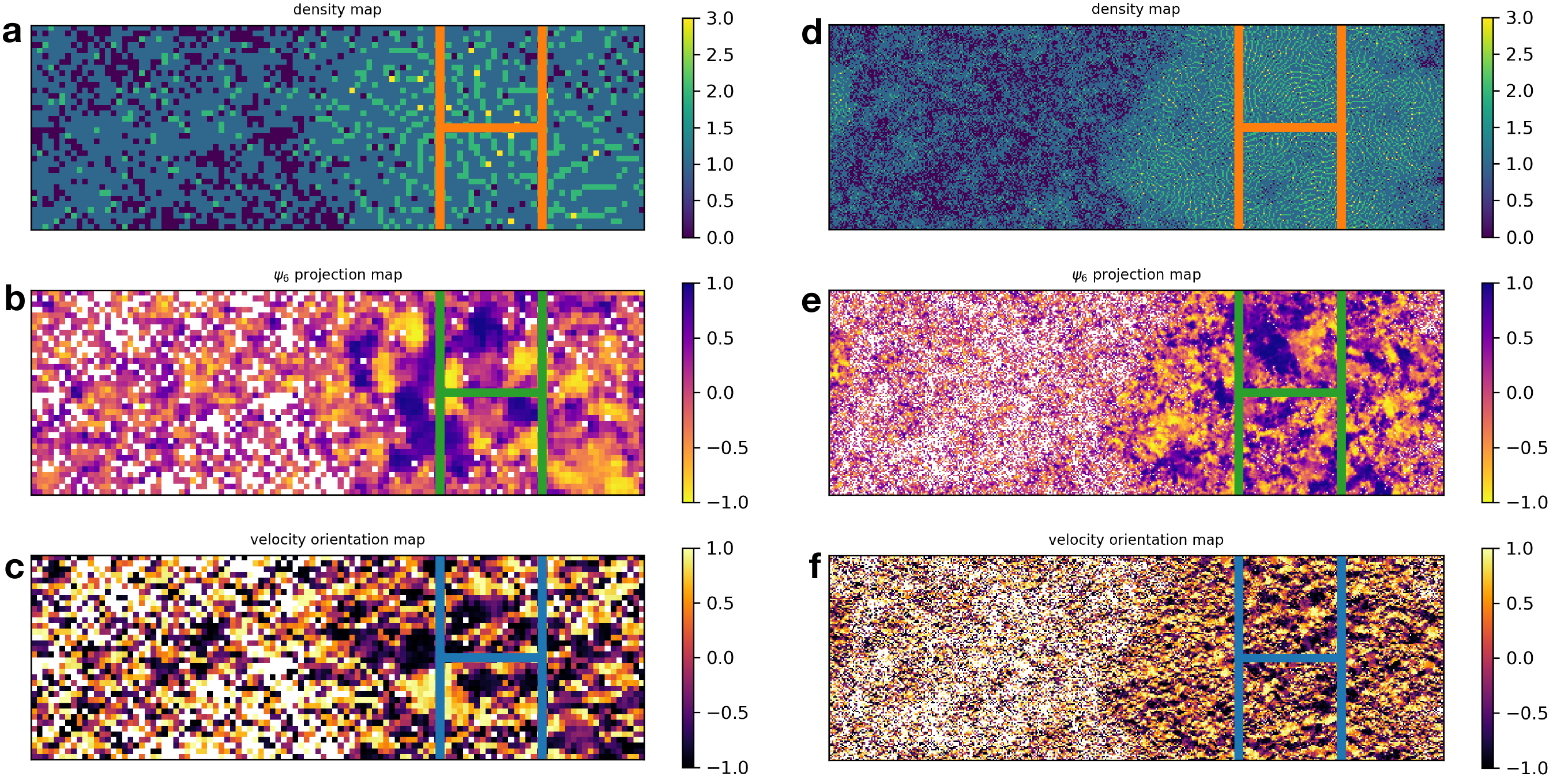}
\caption{(a), (b) and (c) represent, respectively the maps of the density field, the $\psi_6$ projection and the velocity direction projection for a typical configuration of sytem with $N=3750$, $\tau=16.5$ and $\rho=0.95$. The map is calculated for a single configuration choosing bins of the order of the particles size. Different colors represent different values of the fields (see color-bars on the right). White pixels in (b) and (c) correspond to bins where no particles are found. The dense-phase sub-boxes (employed for the FSS) are drawn on (a), (b) and (c) to compare its size with the size of hexaitc andvelocity-oriented domains. (d), (e) and (f) are the same of (a),(b) and (c) respectively but for a configuration of a large system ($N=60 \times 10^3$, $\tau=16.5$ and $\rho=0.95$). }
\label{fig:vel_corr}
\end{figure*}

\begin{figure*}[!]
\includegraphics
[width=0.4\textwidth]{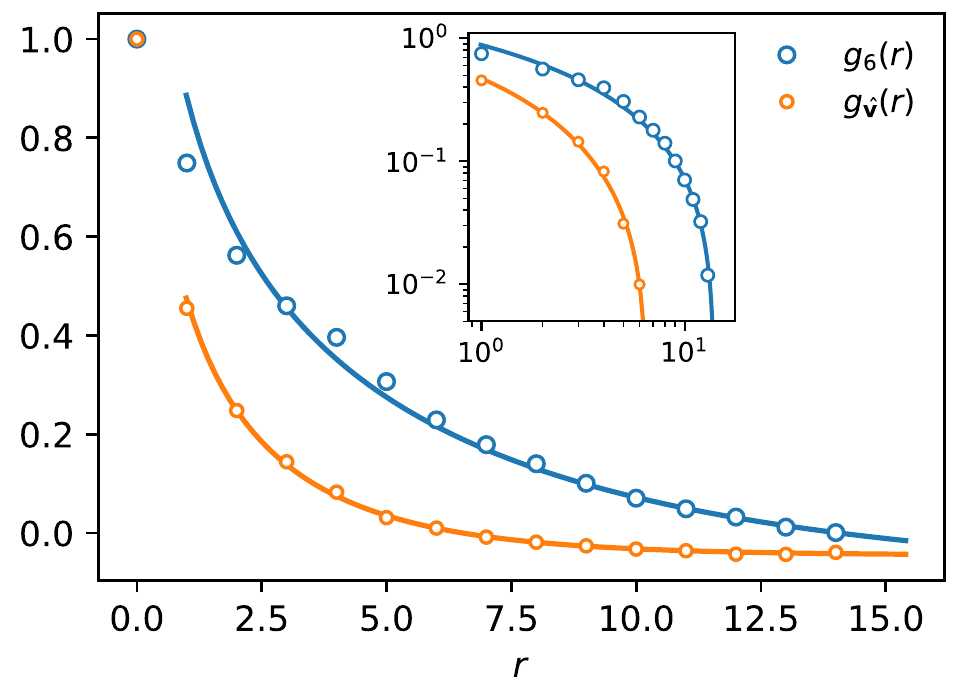}
\caption{Spatial correlation function of the hexatic order parameter $\psi_6$ (blue points) and of the velocity orientation vector (orange points, see legend) for particles in the dense phase for the system with ($N=60 \times 10^3$, $\tau=16.5$ and $\rho=0.95$).
The full lines are fits with the $K_0(r/\xi)$ Bessel function.
The inset is the same of the main panel but on a double-log scale.
}
\label{fig:g6}
\end{figure*}

 {A very recent work~\cite{caporusso2020micro} has shown that the dense phase formed by active particles undergoing MIPS is made of a mosaic of hexatic micro-domains whose size does not diverge.
To compare the size of these regions with our smallest system size, near the critical point, we consider the state point $\tau=16.5$, $\rho=0.95$ for $N=3750$.
In Fig.~\ref{fig:vel_corr}(a) we show a high-resolution density map of one configuration of this system (near $\tau_c$, already showing phase separation). This $\rho$-map is obtained by counting the number of particles in small squared bins of linear size $s=1$.
To characterize the hexatic order we calculate the  parameter $\psi_{6j}=N_j^{-1}\sum_{k}e^{\mathrm{i} \theta_{jk}}$ for each particle. Here $\theta_{jk}$ is orientation angle of the segment connecting the position of the $j$-th particle with its
$k$-th (out of $N_j$) nearest neighbors found with a Voronoi tessellation. 
To visualize the regions with the same orientation we project $\psi_{6j}$ onto the direction of the mean orientation ${N^{-1}\sum_i \psi_{6i}}$ where the sum runs over all particles in the system.
In Fig.~\ref{fig:vel_corr}(b) we show the $\psi_6$-projection map obtained by averaging the $\psi_6$-projection of the particles found in each small bin (white pixels correspond to empty bins). In Fig.~\ref{fig:vel_corr}(b) it is evident that, in the dense phase, hexatic domains (i.e. regions with the same color) have an extent comparable to the size $L$ of the FSS analysis boxes (we have $L\approx 18$ for $N=3750$).\\
\indent
\\
\indent Recent works~\cite{caprini2020spontaneous,caprini2020hidden} have also shown that in active systems the colored noise induces an effective coupling between particles velocities. This effect gives rise to regions of densely packed particles with correlated speed and velocity orientation. We show here that, close to $\tau_c$, these regions have a size similar to the one of the hexatic regions. To visualize the extent of these velocity correlations we show in Fig.~\ref{fig:vel_corr}(c) the orientation map of particle velocities.
This map is obtained by averaging the projected particle velocity vector on the $x$-axis, i.e. $\cos(\vartheta_j)$ (where $\vartheta_j$ is the orientation angle of the $j$-th particle velocity).
Fig.~\ref{fig:vel_corr}(c) shows that the ``islands'' of velocity-correlated particles have a size comparable with the size of hexatic regions. Note however that when we consider a larger system ($N=60 \times 10^3$ and $L\approx72$) at the same $\tau$ and $\rho$ that is phase separating (Fig.~\ref{fig:vel_corr}(d)) the extension of these correlated hexatic and velocity regions does not scale up but remains approximately of the same size (see Fig.~\ref{fig:vel_corr}(e) and (f)). To quantify this more precisely we compute the correlation function of the hexatic order parameter $g_6(r)=\langle \psi^{\ast}_{6j} \psi_{6k}\rangle_{|\mathbf{r}_k-\mathbf{r}_j|=r}/\langle |\psi_{6j}|^2  \rangle$ and the correlation function of the velocity orientation vector
$g_{\hat{\mathbf{v}}}(r)=\langle \hat{\mathbf{v}}_j \cdot \hat{\mathbf{v}}_k \rangle_{|\mathbf{r}_k-\mathbf{r}_j|=r}$.
These functions are computed and reported in Fig.~\ref{fig:g6} considering only particles in the dense phase of the largest system. We find that both $g_6$ and $g_{\hat{\mathbf{v}}}$ decay to zero in an exponential-like fashion as shown in the double-log inset Fig.~\ref{fig:g6}. We assume that both correlators are well described by a Ornstein-Zernike form in $q$-space (i.e. $g(q) \sim (\xi^{-2} + q^2 )^{-1}$) and therefore we fit both data-sets with a function of the form 
$g(r)=A\,K_0(r/\xi)+B$ where $K_0$ is the modified Bessel function of the second kind $\xi$ is the correlation length and $A$ and $B$ are amplitude and shift factors. The fit is quite good and reveals (in agreement with the qualitative map analysis discussed above) that the typical correlation lengths of the hexatic domains and velocity-oriented domains are, respectively, $\xi_6=9.9(1.2)$ and $\xi_{\hat{\mathbf{v}}} = 4.89(0.15)$.}
\indent

\end{document}